\documentclass[journal=jctcce,manuscript=article,layout=twocolumn]{achemso}

\usepackage[colorlinks,linkcolor=blue,citecolor=blue,urlcolor=black,bookmarks=false,hypertexnames=true]{hyperref} 
\usepackage[version=3]{mhchem}
\usepackage{amssymb}
\usepackage{amsmath}
\usepackage{color}
\usepackage{braket}
\usepackage{xspace}
\usepackage{cleveref}
\usepackage{graphicx}
\usepackage{subfigure}
\usepackage{threeparttable}
\usepackage{textcomp}
\usepackage{setspace}
\usepackage{caption}
\usepackage{comment}
\usepackage{cancel}
\usepackage{booktabs, multirow} % for borders and merged ranges
\usepackage{soul}% for underlines
\usepackage[table]{xcolor} % for cell colors
\usepackage{changepage,threeparttable} % for wide tables

\usepackage{array}
\newcolumntype{L}[1]{>{\raggedright\let\newline\\\arraybackslash\hspace{0pt}}m{#1}}
\newcolumntype{C}[1]{>{\centering\let\newline\\\arraybackslash\hspace{0pt}}m{#1}}
\newcolumntype{R}[1]{>{\raggedleft\let\newline\\\arraybackslash\hspace{0pt}}m{#1}}

\newcommand{\angstrom}{\mbox{\normalfont\AA}\xspace}

\DeclareUnicodeCharacter{2009}{\,}

\crefname{figure}{Figure}{Figures}
\crefname{table}{Table}{Tables}
\crefname{equation}{Eq.}{Eqs.}
\crefname{section}{Section}{Sections}
\crefname{subsection}{Section}{Sections}

\makeatletter
\let\l@addto@macro\relax
\makeatother
\usepackage[fontsize=11pt]{scrextend}

\let\oldmaketitle\maketitle
\let\maketitle\relax
\author{James~D.~Serna$^\dag$}
\affiliation{$^\dag$Department of Chemistry and Biochemistry, The Ohio State University, Columbus, Ohio 43210, USA}

\author{Alexander~Yu.~Sokolov$^\dag$}
\email{sokolov.8@osu.edu}
\affiliation{$^\dag$Department of Chemistry and Biochemistry, The Ohio State University, Columbus, Ohio 43210, USA}

\begin{tocentry}
\includegraphics[width=1.0\textwidth]{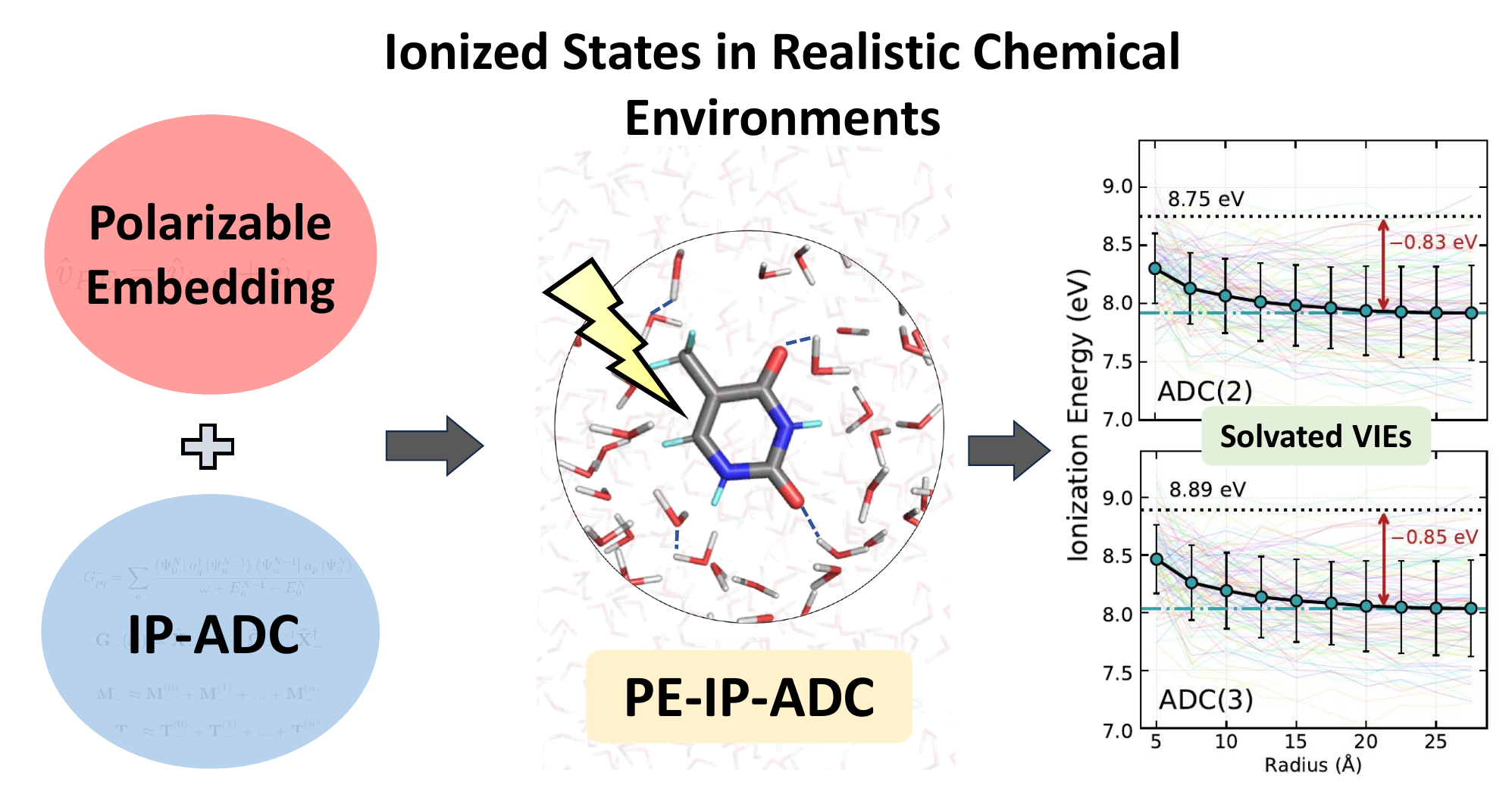}
\end{tocentry}

\title{{\color{blue}
    Simulating Ionized States in Realistic Chemical Environments With Algebraic Diagrammatic Construction Theory and Polarizable Embedding
}}

\begin{document}

%\titlepage

%ArXiv %%%%%%%%%%%%%%
\newcommand*{\abstractext}{
Theoretical simulations of electron detachment processes are vital for understanding chemical redox reactions, semiconductor and electrochemical properties, and high-energy radiation damage. 
However, accurate calculations of ionized electronic states are very challenging due to their open-shell nature, importance of electron correlation effects, and strong interactions with chemical environment. 
In this work, we present an efficient approach based on algebraic diagrammatic construction theory with polarizable embedding that allows to accurately simulate ionized electronic states in condensed-phase or biochemical environments (PE-IP-ADC).
We showcase the capabilities of PE-IP-ADC by computing the vertical ionization energy (VIE) of thymine molecule solvated in bulk water.
Our results show that the second- and third-order PE-IP-ADC methods combined with the basis of set of triple-zeta quality yield a solvent-induced shift in VIE of $-$0.92 and $-$0.93 eV, respectively, in an excellent agreement with experimental estimate of $-$0.9 eV.
This work demonstrates the power of PE-IP-ADC approach for simulating charged electronic states in realistic chemical environments and motivates its further development.
\vspace{0.25cm}
}
%ArXiv %%%%%%%%%%%%%%

%ArXiv %%%%%%%%%%%%%%
\twocolumn[
\begin{@twocolumnfalse}
\oldmaketitle
\vspace{-0.75cm}
\begin{abstract}
\abstractext
\end{abstract}
\end{@twocolumnfalse}
]
%ArXiv %%%%%%%%%%%%%%

%\maketitle
\section{Introduction}
\label{sec:introduction}
Simulating electron detachment processes is crucially important for understanding a variety of phenomena, such as radiation damage of biomolecules,\cite{murray1992271-281a,reisz2014260-292,riley199427-33}  conductivity in materials,\cite{diarra2007045301, graham20161601211,kotadiya2018329-334} and redox reactions in condensed phases or at interfaces.\cite{bao201014467-14479,sauers2010602-606,leopoldini20044916-4922}  
However, accurate modeling of ionized states using electronic structure methods has proven to be challenging.\cite{mckechnie2015194114} 
The sudden removal of an electron can cause significant perturbations to chemical properties, requiring accurate depictions of electron correlation, orbital relaxation, and open-shell electronic states. 
Moreover, in condensed phases, electron detachment results in a localized charged state\cite{banerjee20233037-3053, politzer2007119-137a}  that strongly interacts with its chemical environment by means of electrostatic and non-covalent forces.\cite{ackerman2018109-128,khamis20212499-2513,hurvois20051829-1839,topham2011165405,kim19896055-6060,agren1991425-467, canuto2010214507} 

A common approach to simulate environmental effects is to employ a dielectric continuum model, such as PCM\cite{mennucci2012386-404a,amovilli1998227-261,herbert2021e1519,scalmani2010114110,mennucci2012386-404, dolgounitcheva20123840-3848} or COSMO.\cite{herbert2021e1519, klamt2011699-709}
Although the continuum models have proven to be extremely useful and are computationally very efficient, they are less accurate for simulating charged electronic states that are strongly interacting with the environment.
A more realistic approach is to combine the quantum mechanical description of a localized charged state with the classical (molecular mechanics) treatment of its surrounding (QM/MM).\cite{boulanger201872-76, senn20091198-1229} 
The QM/MM method of choice must be efficient and accurate to: 
(i) describe the charge distribution in a sufficiently large QM region, 
(ii) capture the electronic structure of the resulting open-shell state including the effects of electron correlation and orbital relaxation,
(iii) incorporate the strong polarization of MM environment as a result of the sudden changes in the QM electronic density,
and (iv) enable sampling over many nuclear configurations during the QM/MM dynamics. 

Many quantum mechanical approaches have been adapted for the QM/MM simulations of charged electronic states, including ground-state and time-dependent density functional theory,\cite{ferre2016density, laurent20132019-2039, morzan2014164105, diazmiron20209503-9512, sulpizi2005671-682} linear-response and equation-of-motion coupled cluster theory\cite{sekino1984255-265, kowalski20041715-1738, ghosh20116028-6038, caricato20133035-3042, slipchenko20108824-8830, mukherjee2020e26127, ranga20201019-1027}, second-order approximate coupled cluster singles and doubles\cite{christiansen1995409-418, osted20032055-2071, olsen2010249-256}, and Green’s function with screened Coulomb interaction \cite{conte20091822-1828, bagheri20161743-1756, li2018035108}.
To incorporate the environment polarization effects, several techniques have been developed, namely: polarizable MM force fields,\cite{halgren2001236-242, baker2015241-254, borodin200911463-11478, boulanger20141795-1809, kratz20161019-1029} effective fragment potential (EFP),\cite{gordon2007177-193,day19961968-1986,ghosh201012739-12754} polarizable embedding (PE)\cite{olsen2011107-143}
and its modifications.\cite{olsen20155344-5355,vandenheuvel20233248-3256} 
Both EFP and PE represent the environment using classical potentials parameterized from ab initio calculations.
An attractive feature of the PE framework is the ability to treat QM-MM polarization effects self-consistently, which may be important when the QM and MM regions strongly interact with each other. 

In this work, we present an approach for simulating ionized states in realistic chemical environments based on the combination of PE with algebraic diagrammatic construction theory (ADC).\cite{banerjee20233037-3053,dreuw20236635-6646,schirmer19984734-4744,trofimov2005144115,dempwolff2019064108,dempwolff2020024113,dempwolff2020024125,banerjee2019224112,banerjee2021074105}
The ADC methods efficiently incorporate electronic correlation and orbital relaxation effects, allowing to perform accurate calculations of ionized states in chemical systems with more than 1000 molecular orbitals.
A combination of PE with ADC (PE-ADC) for neutral electronically excited states has been reported.\cite{scheurer20196154-6163, scheurer20184870-4883}
Here, we extend the PE-ADC approach to directly compute the electron-detached states in realistic environments (PE-IP-ADC). 

This paper is organized as follows. 
First, we briefly review the ADC and PE approaches, and discuss the formulation of PE-ADC for ionized states (\cref{sec:theory}).
Following the presentation of computational details (\cref{sec:comp}), we verify the correctness of our PE-IP-ADC implementation against the results from PE-ADC for neutral excitations (\cref{sec:results:validation}).
In \cref{sec:results:benchmark}, we benchmark the accuracy of polarizable embedding for the thymine molecule solvated in small water clusters. 
Finally, we demonstrate the PE-IP-ADC capabilities by calculating the vertical ionization energy of thymine in bulk water (\cref{sec:results:solvated_thymine}). 
Our conclusions and perspectives for future work are presented in \cref{sec:conclusions}.

\section{Methods}
\label{sec:theory}

\subsection{Algebraic Diagrammatic Construction Theory}
\label{sec:theory:adc}

We begin with a brief review of algebraic diagrammatic construction theory for ionized states (IP-ADC).\cite{banerjee20233037-3053,dreuw20236635-6646,schirmer19984734-4744,trofimov2005144115,dempwolff2019064108,dempwolff2020024113,dempwolff2020024125,banerjee2019224112,banerjee2021074105}
IP-ADC uses low-order perturbation theory to approximate the backward component of the one-particle Green's function (1-GF), which in its spectral (or Lehmann) representation can be written as:
\begin{align}
	\label{eq:ipgf}
	G_{-pq} = \sum_n \frac{\bra{\Psi_0^N} a_q^{\dagger} \ket{\Psi_n^{N-1}} \bra{\Psi_n^{N-1}} a_p \ket{\Psi_0^N}}{\omega + E_n^{N-1} - E_0^N}
\end{align}
Here, $\ket{\Psi_0^N}$ is the exact ground-state $N$-electron wavefunction with energy $E_0^N$, while $\ket{\Psi_n^{N-1}}$ and $E_n^{N-1}$ are the exact wavefunctions and energies of the ionized system with $N-1$ electrons. 
The denominator of 1-GF depends on the ionization energies ($\tilde{\Omega}_{-n} = E_0^N - E_n^{N-1}$). 
The numerator is written in terms of spectroscopic amplitudes ($\tilde{X}_{-qn} = \bra{\Psi_0^N} a_q^{\dagger} \ket{\Psi_n^{N-1}}$) that describe the probability of electron detachment in photoelectron spectra and the  creation and annihilation operators ($a_p^\dagger$ and $a_p$).
\cref{eq:ipgf} can be written more compactly in a matrix form
\begin{align} 
	\label{eq:2}
	\mathbf{G}_{-}(\omega) = \tilde{\mathbf{X}}_{-}(\omega \mathbf{1} - \tilde{\mathbf{\Omega}}_{-})^{-1}\tilde{\mathbf{X}}^{\dagger}_{-}
\end{align}
where $\tilde{\mathbf{\Omega}}_{-}$ is a diagonal matrix of ionization energies ($\tilde{\Omega}_{-n}$).

Since the exact eigenstates and their energies are usually not known, IP-ADC expresses 1-GF in a non-diagonal matrix form
\begin{align}
	\label{eq:adc_1gf}
	\mathbf{G}_{-}(\omega) = \mathbf{T}_{-}(\omega \mathbf{1}- \mathbf{M}_{-})^{-1} \mathbf{T}_{-}^{\dagger}
\end{align}
where matrix elements are evaluated in an orthonormal basis set of $(N-1)$-electron configurations. 
Similar to $\tilde{\mathbf{\Omega}}_{-}$ and $\tilde{\mathbf{X}}_{-}$ in \cref{eq:2}, $\mathbf{M}_{-}$ (``effective Hamiltonian'' matrix) and $\mathbf{T}_{-}$ (``effective transition moments'' matrix) in \cref{eq:adc_1gf} contain information about the ionization energies and photoelectron probabilities, respectively. 
Each matrix is evaluated up to the order $n$ in perturbation theory
\begin{align}
	\mathbf{M}_- &\approx \mathbf{M}_-^{(0)} + \mathbf{M}_-^{(1)} + ... + \mathbf{M}_-^{(n)} \\ 
	\mathbf{T}_- &\approx \mathbf{T}_-^{(0)} + \mathbf{T}_-^{(1)} + ... + \mathbf{T}_-^{(n)}  
\end{align}
defining the $n$th-order IP-ADC approximation (IP-ADC($n$)). 
The approximate ionization energies (IP's) are computed by diagonalizing $\mathbf{M}_-$:
\begin{align}
	\label{eq:diag}
	\mathbf{M}_- \mathbf{Y}_- = \mathbf{Y}_- \boldsymbol{\Omega}_-
\end{align}
Combining $\mathbf{T}_-$ with the eigenvectors $\mathbf{Y}_-$ allows to calculate the approximate spectroscopic amplitudes
\begin{align}
	\mathbf{X}_- = \mathbf{T}_- \mathbf{Y}_-
\end{align}
which can be used to evaluate the photoelectron probabilites (so-called spectroscopic factors):
\begin{align}
	\label{eq:spec_factors}
	P_{-n} = \sum_{p} |X_{- pn}|^{2}
\end{align}
The IP-ADC($n$) methods can be formulated using single-reference\cite{banerjee20233037-3053,banerjee20225337-5348,dreuw20236635-6646,schirmer19984734-4744,trofimov2005144115,dempwolff2019064108,dempwolff2020024113,dempwolff2020024125,banerjee2019224112,banerjee2021074105} or multireference perturbation theory.\cite{sokolov2018204113a,chatterjee20195908-5924,chatterjee20206343-6357,mazin20216152-6165}
In this work, we will focus on the single-reference IP-ADC approximations.

We note that the IP-ADC approach presented here is referred to as the non-Dyson approach in the ADC literature.\cite{schirmer19984734-4744,trofimov2005144115,dempwolff2019064108,dempwolff2020024113,dempwolff2020024125,banerjee2019224112,banerjee2021074105, banerjee20233037-3053}
An alternative formulation of ADC based on the Dyson equation was also developed,\cite{schirmer19831237-1259} but will not be employed in this work. 

\subsection{Polarizable Embedding}
\label{sec:theory:pe}

We now briefly discuss the framework of polarizable embedding (PE).\cite{olsen2011107-143, olsen20103721-3734, steinmann2019e25717, bondanza202014433-14448, list201620234-20250} In PE, the chemical system is split into a quantum mechanical (QM) region and a classical polarizable environment with energy
\begin{align}
	\label{eq:pe_tot}
	E_{\mathrm{tot}} = E_{\mathrm{QM}} + E_{\mathrm{env}} + E_{\mathrm{es}}^{\mathrm{PE}} + E_{\mathrm{ind}}^{\mathrm{PE}}
\end{align}
where $E_{\mathrm{QM}}$ is the total energy of QM region, $E_{\mathrm{env}}$ is the environment energy, and the remaining two terms are the electrostatic ($E_{\mathrm{es}}^{\mathrm{PE}} $) and induction ($ E_{\mathrm{ind}}^{\mathrm{PE}}$) energies of QM--environment interaction. 
The environment region is further separated into individual fragments (sites) corresponding to either solvent molecules or fragments of larger (bio-) molecules that are described using nonempirically parametrized multipoles and polarizabilities.

The electrostatic contribution to the QM--environment interaction energy incorporates the interactions between the nuclei and electrons with the permanent multipoles representing the environment.
It can be expressed as\cite{olsen2011107-143, olsen20103721-3734, steinmann2019e25717, bondanza202014433-14448, list201620234-20250} 
\begin{align}
	\label{eq:energy_es}
	E_{\mathrm{es}}^{\mathrm{PE}} = \sum_{s=1}^S \sum_{k=0}^K \frac{(-1)^{k}}{k!}
	\left(\sum_{m=1}^M Z_m \mathbf{W}_{ms}^{(k)} - \sum_{i=1}^N \mathbf{W}_{is}^{(k)}\right) \mathbf{Q}_s^{(k)} 
\end{align}
where $\mathbf{Q}_s^{(k)}$ is the $k$th-order multipole moment for the $s$th site in the environment, $Z_m$ is the nuclear charge of the $m$th nucleus, and $\mathbf{W}^{(k)}$ are the interaction tensors for the nuclei ($\mathbf{W}_{ms}^{(k)}$) and electrons ($\mathbf{W}_{is}^{(k)}$).\cite{steinmann2019e25717}
The values $S$, $M$, and $N$ represent the number of environment sites, nuclei, and electrons, respectively.
For each site, the multipole expansion is truncated at order $K$.

To describe the induction effects, each environment site is represented with an induced dipole that interacts with electric field (${\mathbf{F}}$) giving rise to the energy contribution
\begin{align}
	\label{eq:energy_ind}
	E_{\mathrm{ind}}^{\mathrm{PE}} 
	= -\frac{1}{2} {\boldsymbol{\mu}}^{\mathrm{ind}} \cdot {\mathbf{F}}
	= -\frac{1}{2} {\boldsymbol{\mu}}^{\mathrm{ind}} \cdot ({\mathbf{F}}^{\mathrm{nuc}} + {\mathbf{F}}^{\mathrm{elec}} + {\mathbf{F}}^{\mathrm{es}})
\end{align}
where ${\boldsymbol{\mu}}^{\mathrm{ind}}$ is a 3$S$-dimensional vector that contains the full set of induced dipoles. 
The electric fields originating from the nuclei, electrons, and permanent multipoles in the environment are denoted as ${\mathbf{F}}^{\mathrm{nuc}}$, ${\mathbf{F}}^{\mathrm{elec}}$, and ${\mathbf{F}}^{\mathrm{es}}$, respectively. 
The induced dipoles are determined as
\begin{align}
	\label{eq:mu_ind}
	{\boldsymbol{\mu}}^{\mathrm{ind}}= \mathbf{B}{\mathbf{F}}
\end{align}
where $\mathbf{B}$ is the symmetric (3$S$ $\times$ 3$S$) classical response matrix 
\begin{align}
	\label{eq:resp_mat}
	\mathbf{B} = 
	\begin{pmatrix}
		\boldsymbol{\alpha}_1^{-1} & \mathbf{W}_{12}^{(2)} & \ldots & \mathbf{W}_{1S}^{(2)} \\
		\mathbf{W}_{21}^{(2)} & \boldsymbol{\alpha}_2^{-1} & \ddots & \vdots \\
		\vdots & \ddots & \ddots & \mathbf{W}^{(2)}_{(S-1)S} \\
		\mathbf{W}^{(2)}_{S1} & \hdots & \mathbf{W}_{S(S-1)}^{(2)} & \boldsymbol{\alpha}_S^{-1}
	\end{pmatrix}^{-1}
\end{align}
with the inverse polarizability tensors on the diagonal ($\boldsymbol{\alpha}_s^{-1}$) and the off-diagonal dipole-dipole interaction tensors ($\mathbf{W}_{ss'}^{(2)}$).

\subsection{Combining PE with ADC}
\label{sec:theory:pe_adc}
In this work, we combine polarizable embedding (PE) with IP-ADC to incorporate the environment polarization effects in the calculations of ionization energies.
A combination of PE with ADC for neutral (particle-number-conserving) excitations has been previously reported (PE-ADC).\cite{scheurer20184870-4883}
Here, we extend this approach to calculations of ionization energies (PE-IP-ADC).

In PE-ADC, the interaction with environment is incorporated in three steps.
First, PE is included in the reference self-consistent field (SCF) calculation to compute the ground-state spin-orbitals of the QM region in the presence of environment (PE-SCF). 
These spin-orbitals are calculated as eigenfunctions of effective one-electron operator
\begin{align}
	\label{eq:total_f_pe}
	\hat{f}_{\mathrm{eff}} = \hat{f}_{\mathrm{HF}} + \hat{v}_{\mathrm{PE}}
\end{align}
where $\hat{f}_{\mathrm{HF}}$ is the standard Fock operator and $\hat{v}_{\mathrm{PE}}$ is the PE potential, which is derived by minimizing the energy functional in \cref{eq:pe_tot} with respect to electronic density variations.
The PE potential operator has the form:
\begin{align}
	\hat{v}_{\mathrm{PE}} &= \hat{v}_{\mathrm{es}} + \hat{v}_{\mathrm{ind}} \\
	\label{eq:ele_oper}
	\hat{v}_{\mathrm{es}} &= \sum_{s=1}^S \sum_{k=0}^K \frac{(-1)^{(k+1)}}{k!} \mathbf{Q}_s^{(k)} \sum_{pq} \mathbf{W}_{pq,s}^{(k)} a^{\dagger}_{p} a_{q} \\
	\label{eq:pe_ind}
	\hat{v}_{\mathrm{ind}} &= -\sum_{s=1}^S \boldsymbol{\mu}_{s}^{\mathrm{ind}}\sum_{pq} \mathbf{W}_{pq,s}^{(1)} a^{\dagger}_{p} a_{q}
\end{align}
where $\mathbf{W}_{pq,s}^{(k)}$ are the interaction tensors evaluated in the basis of spin-orbitals with indices $p$ and $q$.
In \cref{eq:pe_ind}, the induced dipole moments ($\boldsymbol{\mu}_{s}^{\mathrm{ind}}$) depend on the total electric field ($\mathbf{F}$) at each site (\cref{eq:mu_ind}), which is calculated from the electronic density in the QM region. 
For this reason, both $\hat{f}_{\mathrm{HF}}$ and $\hat{v}_{\mathrm{PE}}$ in \cref{eq:total_f_pe} are updated at each SCF iteration until full self-consistency.
At convergence, $\hat{f}_{\mathrm{eff}}$ is fully diagonal in the basis of its eigenfunctions, with the eigenvalues representing the energies of spin-orbitals for a polarized system.

In the second step, the excitation energies of QM region are computed using ADC starting with the PE-SCF molecular orbitals and orbital energies corresponding to the eigenfunctions and eigenvalues of $\hat{f}_{\mathrm{eff}}$.
This calculation incorporates the Coulomb interaction of ground and excited states with the electrostatic environment, but misses the polarization effects due to the changes in electronic density upon excitation.

In the final step, the ADC excitation energies are supplied with perturbative corrections ($\Delta E^{\mathrm{PE,corr}}_{0\rightarrow n}$) that account for the missing polarization effects:
\begin{align}
	\label{eq:energ_exc_tot}
	\Delta E_{0 \rightarrow n}^{\mathrm{PE\mbox{-}ADC}} 
	&= \Delta E_{0 \rightarrow n}^{\mathrm{ADC}} 
	+ \Delta E^{\mathrm{PE,corr}}_{0\rightarrow n}
\end{align}
As discussed in Ref.\@ \citenum{scheurer20184870-4883}, $\Delta E^{\mathrm{PE,corr}}_{0\rightarrow n}$ can be approximated as:
\begin{align}
	\label{eq:pert_tot}
	\Delta E^{\mathrm{PE,corr}}_{0\rightarrow n} 
	&= \Delta E^{\mathrm{ptSS}}_{0\rightarrow n} + \Delta E^{\mathrm{ptLR}}_{0\rightarrow n} 
\end{align}
where $\Delta E^{\mathrm{ptSS}}_{0\rightarrow n}$ is a state-specific perturbative correction describing the change in mutual induction energy upon excitation and $\Delta E^{\mathrm{ptLR}}_{0\rightarrow n}$ is a linear-response correction due to the nonresonant excitonic coupling.

To implement PE-IP-ADC, we combined our implementation of non-Dyson IP-ADC in PySCF\cite{sun2020024109} with the CPPE program\cite{scheurer20196154-6163} that allows to calculate the PE-SCF reference wavefunctions and the perturbative corrections to the excitation energies. 
Since electron detachment corresponds to the limit of a dissociated exciton (e.g., cation + electron in the continuum), the $\Delta E^{\mathrm{ptLR}}_{0\rightarrow n}$ correction describing nonresonant excitonic coupling is zero in all IP-ADC calculations and does not need to be computed (see \cref{sec:results:validation} for numerical validation). 
The $\Delta E^{\mathrm{ptSS}}_{0\rightarrow n}$ correction is evaluated as\cite{scheurer20184870-4883}
\begin{align}
	\label{eq:ptss}
	\Delta E^{\mathrm{ptSS}}_{0\rightarrow n} = -\frac{1}{2} {\mathbf{F}}^{\mathrm{elec}} [\Delta \boldsymbol{\gamma}_n]^{\dagger} {\boldsymbol{\mu}}^{\mathrm{ind}}[\Delta \boldsymbol{\gamma}_n]
\end{align}
where ${\mathbf{F}}^{\mathrm{elec}} [\Delta \boldsymbol{\gamma}_n]$ and  ${\boldsymbol{\mu}}^{\mathrm{ind}}[\Delta \boldsymbol{\gamma}_n]$ describe the changes in electric field and induced dipoles originating from the change in electronic density upon ionization ($\Delta \boldsymbol{\gamma}_n = \boldsymbol{\gamma}_n - \boldsymbol{\gamma}_0$), respectively.
In our PE-IP-ADC implementation, this electronic density difference is computed from the ground- and ionized-state reduced density matrices ($\boldsymbol{\gamma}_0$, $\boldsymbol{\gamma}_n$) using the approach outlined in Ref.\@ \citenum{stahl2022044106}.

\subsection{Computational Details} 
\label{sec:comp}

 \begin{figure*}[!t]
	\subfigure[]{
		\includegraphics[width=0.45\textwidth]{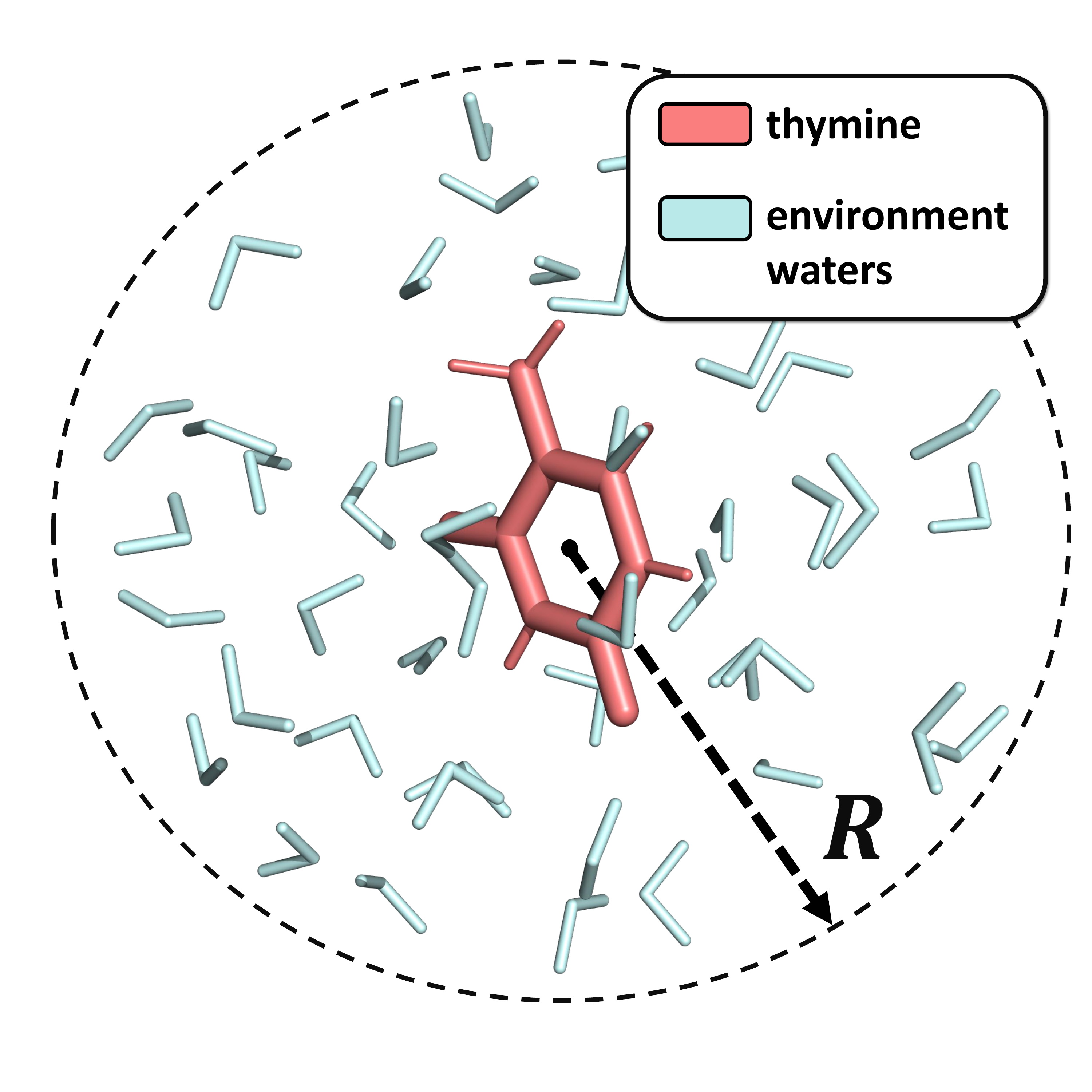} 
		\label{fig:solv_radius_a} }
	\subfigure[]{
		\includegraphics[width=0.45\textwidth]{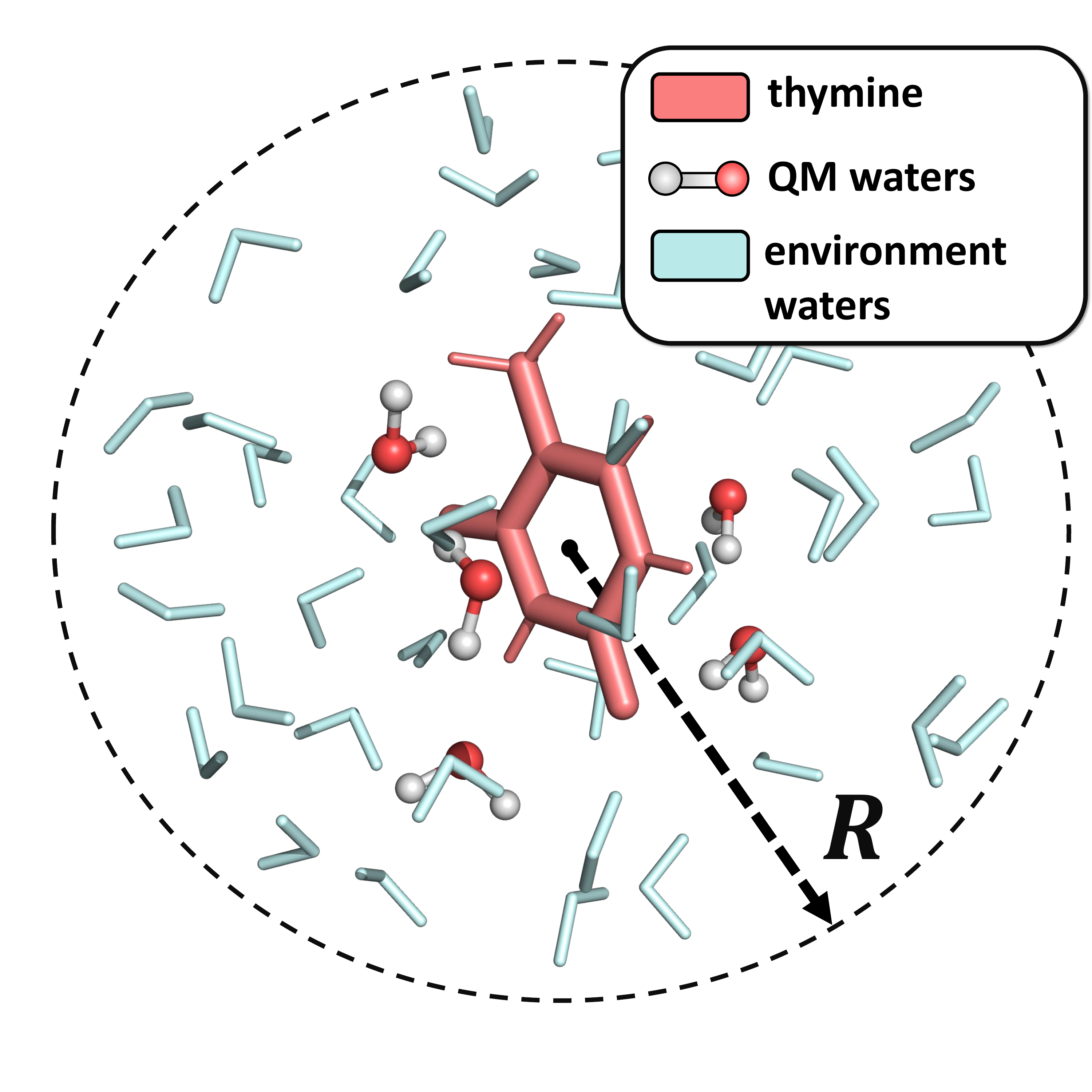}
		\label{fig:solv_radius_b} }
	\captionsetup{justification=raggedright,singlelinecheck=false}
	\caption{
		Two solvated thymine models constructed from the structure in \cref{fig:solv_system} and used for the PE-IP-ADC benchmark studies in \cref{sec:results:benchmark}. 
		In each model, a sphere of radius $R$ is defined relative to the thymine center of mass and the water molecules outside the sphere are discarded. 
		a) Only thymine is included in the QM region.
		b) Thymine and all water molecules within the radius of 4 $\angstrom$ are incorporated in the QM region.
	}
	\label{fig:solv_radius}
\end{figure*}

In \cref{sec:results}, we verify our implementation of PE-IP-ADC methods (\cref{sec:results:validation}), benchmark their accuracy (\cref{sec:results:benchmark}), and use them to compute the ionization energy of thymine molecule in bulk water (\cref{sec:results:solvated_thymine}). 
As discussed in \cref{sec:theory:pe_adc}, the PE-IP-ADC calculations were performed by interfacing the implementation of non-Dyson IP-ADC in PySCF\cite{sun2020024109} with the CPPE program.\cite{scheurer20196154-6163} 
CPPE requires input of an embedding potential that is generated using PyFraME\cite{steinmann2019e25717} and Dalton\cite{aidas2014269-284} for a user-defined environment surrounding the QM region. \cite{list201620234-20250}

To validate our PE-IP-ADC implementation (\cref{sec:results:validation}), we compare its results against the calculations using PE-ADC for neutral excitations (PE-EE-ADC)\cite{scheurer20184870-4883} where a very diffuse $s$-function with the exponent of $10^{-10}$ was added to the basis set of oxygen atoms. 
Here, two systems will be investigated: 1) Nile red in $\beta$-lactoglobulin (BLG) and 2) thymine in a small cluster of water. 
The Nile red in BLG calculations were performed using the STO-3G basis set\cite{hehre19692657-2664, pritchard20194814-4820} with the geometry and embedding potential from Ref.\@ \citenum{scheurer20196154-6163}.
The validation tests on thymine were carried out using the cc-pVDZ basis set\cite{dunning19891007-1023, pritchard20194814-4820} for a spherical water cluster of 6 $\angstrom$ radius relative to the thymine center of mass, which included 30 water molecules ({see the Supplementary Information for details}). 
All electrons were correlated in all ADC calculations.

In \cref{sec:results:benchmark}, we benchmark the accuracy of PE-IP-ADC ionization energies against the results from all-quantum IP-ADC calculations for the thymine + water clusters with the solvation shell radius ranging from 3.5 to 7 $\angstrom$.
These tests were performed using the cc-pVDZ, aug-cc-pVDZ, and cc-pVTZ basis sets and density fitting\cite{hattig20005154-5161, hattig200537-60, whitten19734496-4501, vahtras1993514-518, feyereisen1993359-363, dunlap19793396-3402, weigend1997331-340, weigend1998143-152, weigend20024285-4291} the corresponding JKFIT and RI auxiliary basis sets\cite{weigend20023175-3183, weigend20024285-4291} were employed to approximate two-electron integrals for the SCF and ADC methods, respectively. 
Two QM regions were employed in PE-IP-ADC consisting of 1) only the thymine molecule or 2) thymine with water molecules within the 4 $\angstrom$ spherical solvation shell (\cref{fig:solv_radius}). 
The remaining water molecules were described using polarizable embedding (PE).
To compare the performance of PE to that of standard force fields, we also performed IP-ADC calculations with the environment water molecules described using the TIP3P model.\cite{jorgensen1983926-935}

 \begin{figure*}[!t]
	\includegraphics[width=.8\textwidth]{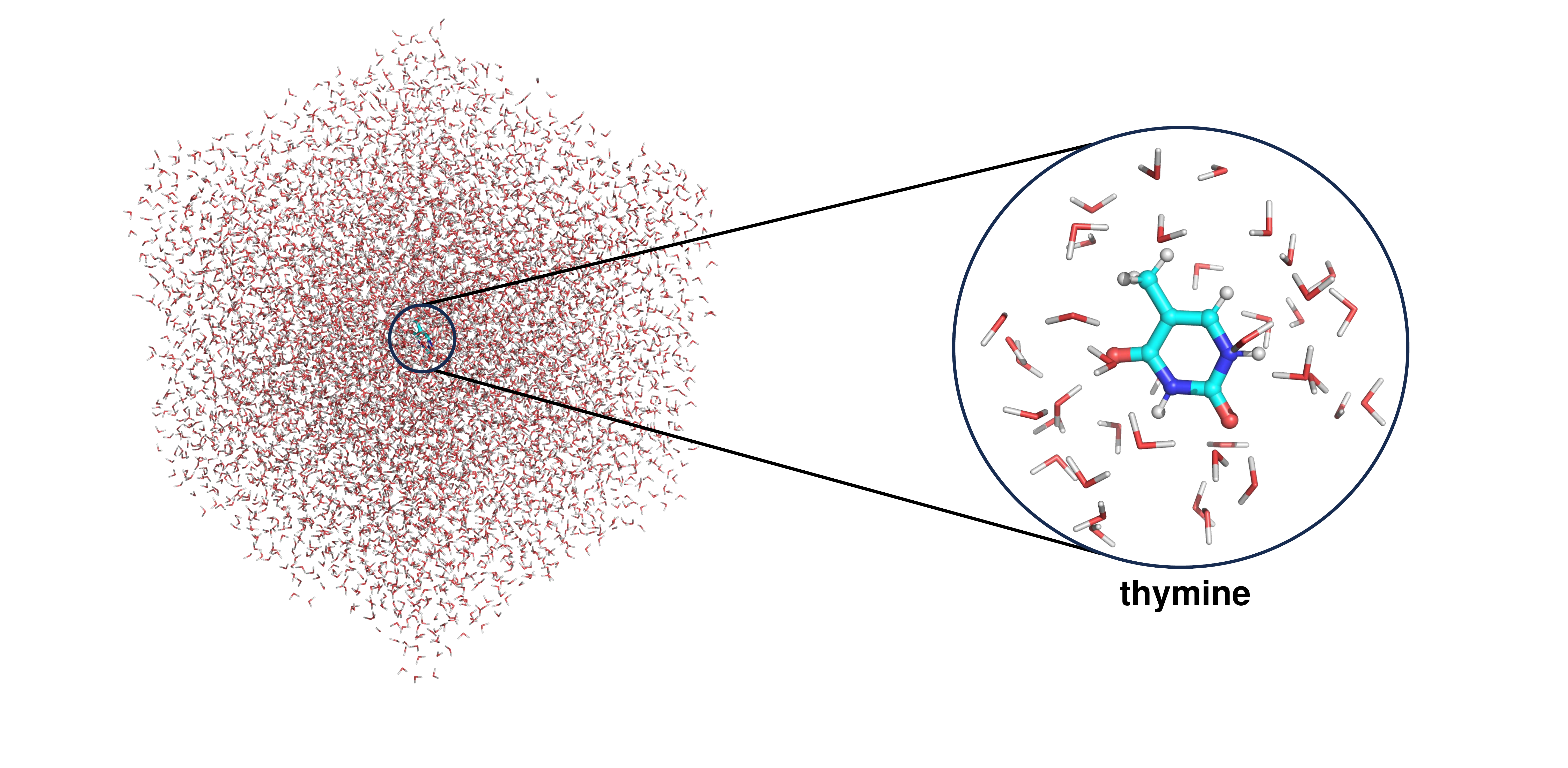}	
	\captionsetup{justification=raggedright,singlelinecheck=false}
	\caption{
		The model of thymine solvated in a $60 \times 60 \times 60\ \angstrom ^3$ box of water used for the molecular dynamics simulations performed in this work.
	}
	\label{fig:solv_system}
\end{figure*}

The solvated thymine structures used in \cref{sec:results:validation,sec:results:benchmark} were computed using the following approach.
First, the geometry of free thymine was optimized using the MP2 level of theory\cite{head-gordon1988503-506} with the polarizable continuum model (PCM)\cite{mennucci2012386-404a,amovilli1998227-261,herbert2021e1519,scalmani2010114110,mennucci2012386-404} and the cc-pVTZ basis set\cite{dunning19891007-1023} implemented in the Psi4 program.\cite{smith2020184108} 
Next, the solvation shell structure was computed by placing the thymine molecule in the center of a $60 \times 60 \times 60\ \angstrom ^3$ box containing 6830 water molecules (\cref{fig:solv_system}).
The positions of solvent molecules were optimized using the nanoscale molecular dynamics (NAMD) interface of VMD\cite{humphrey199633-38} by minimizing the energy of the system for 10000 steps while keeping the thymine geometry frozen. 
The NAMD calculations used the TIP3P model \cite{jorgensen1983926-935} for the water molecules and the CGenFF\cite{vanommeslaeghe2010671-690} force field for thymine. 
To perform the studies in \cref{sec:results:validation,sec:results:benchmark}, we defined a sphere of radius $R$ relative to the thymine center of mass and discarded all water molecules outside of this radius (\cref{fig:solv_radius}).

 \begin{figure*}[!t]
	\includegraphics[width=1\textwidth]{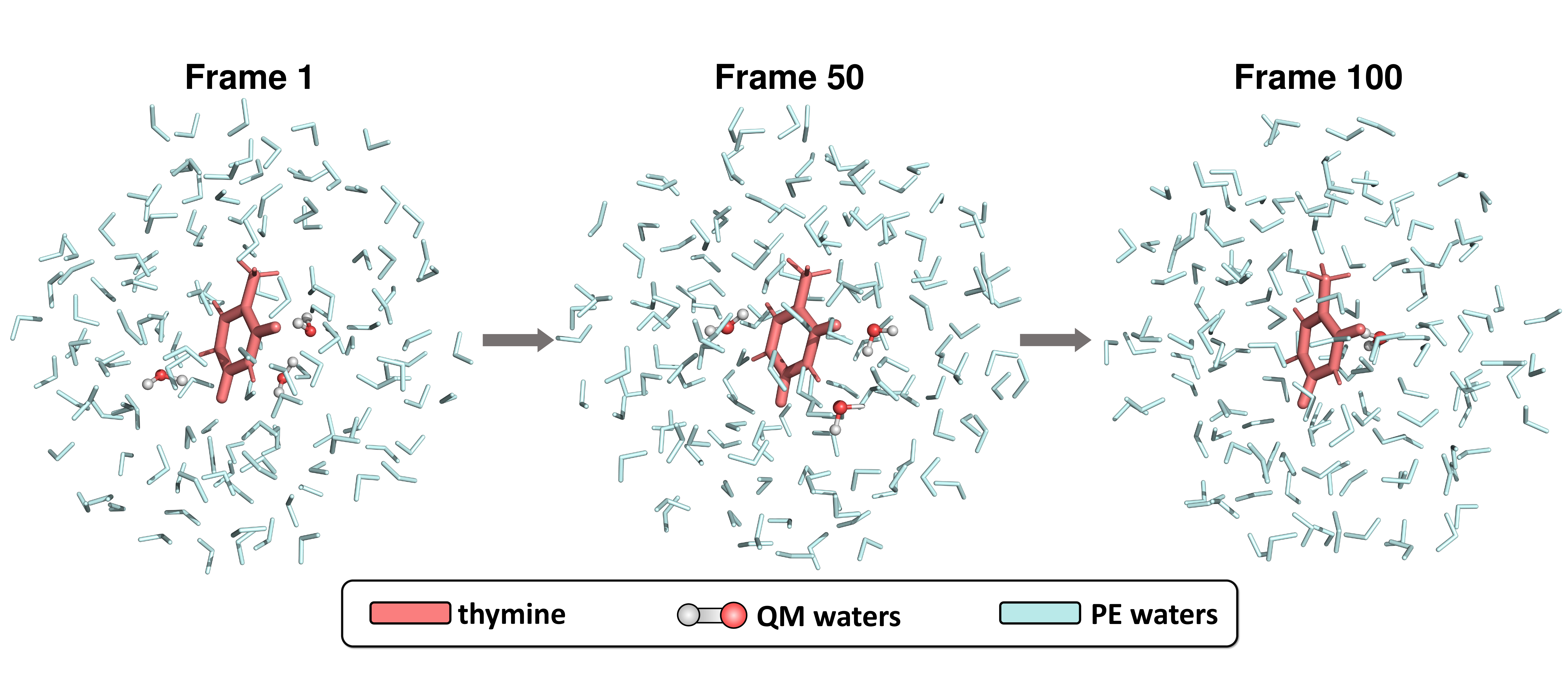}	
	\captionsetup{justification=raggedright,singlelinecheck=false}
	\caption{
		Illustration of sampled geometries from a molecular dynamics simulation of thymine solvated in water. 
		Results are shown for a spherical solvation shell of 10 $\angstrom$ relative to the thymine center of mass. 
		Water molecules in the QM region are visualized using the ball and stick models. 
		The remaining water molecules are described using polarizable embedding. 
	}
	\label{fig:md_frames}
\end{figure*}

\begin{table*}[t!]
	\centering
	\caption{
		Excitation energies ($E$, eV) computed using PE-EE-ADC(2) and PE-IP-ADC(2) for two chemical systems: (i) Nile red in BLG with the STO-3G basis set (structure from Ref.\@ \citenum{scheurer20196154-6163}) and (ii) thymine solvated in water with the cc-pVDZ basis set (see \cref{sec:comp} for details). 
		In the PE-EE-ADC calculations, a very diffuse $s$-orbital was added to the basis set to mimic the ionization continuum.
		Also shown are the state-specific ($\Delta E^{\mathrm{ptSS}}_{0\rightarrow n}$, eV) and linear-response ($\Delta E^{\mathrm{ptLR}}_{0\rightarrow n}$, eV) corrections to the PE-ADC excitation and ionization energies.
		The results for ionized states are highlighted in bold.}
	\centering
	\footnotesize
	\setstretch{1}
	\label{table:ptlr}
	\begin{tabular}{lcccccccccc}
		\hline \midrule
		\multicolumn{5}{c}{{Nile red in BLG}} &  \multicolumn{5}{c}{\hspace{1.2cm}{Thymine in water}} \\ \midrule
		Type &State &$E$ & $\Delta E^{\mathrm{ptSS}}_{0\rightarrow n}$ &$\Delta E^{\mathrm{ptLR}}_{0\rightarrow n}$&  &{State} &$E$ &$\Delta E^{\mathrm{ptSS}}_{0\rightarrow n}$ &$\Delta E^{\mathrm{ptLR}}_{0\rightarrow n}$ \\  \midrule
		EE &{1} &{1.865} &{$-$0.002} &{0} & \hspace{1.2cm} &\textbf{11} &\textbf{8.208} &\textbf{$-$0.223} &\textbf{0} \\
		&\textbf{2} &\textbf{2.194} &\textbf{$-$0.514} &\textbf{0} & & \textbf{12} &\textbf{8.614} &\textbf{$-$0.291} &\textbf{0} \\
		&\textbf{3} &\textbf{2.616} &\textbf{$-$0.671} &\textbf{0} & &13 &8.636 &$-$0.054 &$-$0.001 \\
		&\textbf{4} &\textbf{3.063} &\textbf{$-$0.551} &\textbf{0} & &14 &9.011 &$-$0.019 &$-$0.014 \\	
		&5 &3.119 & $-$0.067 & $-$0.002  & &15 & 9.400 & $-$0.007 & $-$0.002 \\
		&6 & 3.409 & $-$0.014 & 0  & & \textbf{16} &\textbf{9.597} &\textbf{$-$0.271} &\textbf{0} \\
		\midrule
        IP &\textbf{1} &\textbf{2.193} &\textbf{$-$0.515} &\textbf{0} &    &\textbf{1} &\textbf{8.209} &\textbf{$-$0.223} &\textbf{0} \\
		&\textbf{2} &\textbf{2.614} &\textbf{$-$0.671} &\textbf{0} & &\textbf{2} & \textbf{8.614} & \textbf{$-$0.291} & \textbf{0} \\
		&\textbf{3} & \textbf{3.063} & \textbf{$-$0.552} & \textbf{0} & &  \textbf{3} & \textbf{9.597} & \textbf{$-$0.271} & \textbf{0} \\ 
		
		\hline \hline
	\end{tabular}
\end{table*}

In \cref{sec:results:solvated_thymine}, we calculate the ionization energy of thymine in bulk water by performing the PE-IP-ADC calculations along the trajectories from molecular dynamics simulation with periodic boundary conditions.
The initial molecular dynamics calculations were performed by slowly heating the system to 300 K for the first 20 ps starting with the optimized structure for the full $60 \times 60 \times 60\ \angstrom ^3$ box described above (\cref{fig:solv_system}).
The thymine geometry was allowed to fluctuate but the molecule was restricted to remain in the center of the water box. 
Following this, a 10 ns ``production'' molecular dynamics simulation was performed with a time step of 1 fs at constant temperature (300 K) and constant pressure (1 atm).
A total of 100 geometries evenly spaced in time were extracted from this calculation, ensuring statistical independence and a well-sampled space.
The molecular dynamics calculations were performed with NAMD using the TIP3P model \cite{jorgensen1983926-935} for the water molecules and the CGenFF\cite{vanommeslaeghe2010671-690} force field for thymine.
The SHAKE constraints\cite{ryckaert1977327-341} on hydrogen atoms were used during the slow heating and production run steps. 

To investigate the dependence of thymine ionization energy on the size of solvation shell, the PE-IP-ADC calculations were performed by constraining the environment to a sphere with radius ($R$) ranging from 5 to 27.5 $\angstrom$ relative to the thymine center of mass.
All water molecules outside of this sphere were discarded prior to running the PE-IP-ADC calculations.
The QM region was comprised of thymine and all water molecules inside the sphere of 4 $\angstrom$. 
Illustrations of selected MD snapshots, highlighting the QM and PE regions, are provided in \cref{fig:md_frames}.
Four basis sets were employed in this study, namely: cc-pVDZ, cc-pVTZ, aug-cc-pVDZ, and aug-cc-pVTZ.\cite{dunning19891007-1023, kendall19926796-6806}
The ionization energy averaged over 100 sampled geometries was compared to that of free thymine molecule.
The geometry of free thymine was optimized at the MP2/cc-pVTZ level of theory without PCM. 

\section{Results and Discussion}
\label{sec:results}

\subsection{Validation}
\label{sec:results:validation}

We begin by verifying our PE-IP-ADC implementation against the results of PE-ADC for neutral excitations (PE-EE-ADC)\cite{scheurer20184870-4883} computed with a very diffuse $s$-orbital in the basis set (see \cref{sec:comp} for computational details).
Incorporating such basis function allows to mimic the photoelectron continuum providing access to the ionization energies in the PE-EE-ADC calculations. 
\cref{table:ptlr} shows the results of PE-IP-ADC(2) and PE-EE-ADC(2) for two chemical systems: (i) Nile red in BLG and (ii) thymine solvated in water.
In addition to the excitation energies ($E$, eV), the state-specific ($\Delta E^{\mathrm{ptSS}}_{0\rightarrow n}$, eV) and linear-response ($\Delta E^{\mathrm{ptLR}}_{0\rightarrow n}$, eV) perturbative corrections are reported for each electronic state (see \cref{sec:theory:pe_adc} for details).

The ionization energies computed using PE-IP-ADC(2) and PE-EE-ADC(2) are in a very good agreement, deviating by no more than  0.002 eV from each other.
This agreement also holds for the $\Delta E^{\mathrm{ptSS}}_{0\rightarrow n}$ correction due to the change in mutual induction energy, which ranges from $-$0.22 to $-$0.67 eV for the ionized states in \cref{table:ptlr}.
Importantly, the linear-response correction $\Delta E^{\mathrm{ptLR}}_{0\rightarrow n}$ describing the non-resonant excitonic coupling is numerically zero for all ionized states, in agreement with the discussion in \cref{sec:theory:pe_adc}.

Overall, the results in \cref{table:ptlr} confirm the correctness of our PE-IP-ADC implementation.
While PE-EE-ADC can be used to simulate both neutral and ionized states, it is less computationally efficient than PE-IP-ADC due to the larger size of effective Hamiltonian matrix $\mathbf{M}$ that needs to be diagonalized in the ADC eigenvalue problem (\cref{eq:diag}).
This difference in computational cost becomes particularly significant at the third order in perturbation theory where the PE-EE-ADC(3) method exhibits a higher ($\mathcal{O}(N^6)$) computational scaling with the basis set size ($N$) than that of PE-IP-ADC(3) ($\mathcal{O}(N^5)$) for the iterative diagonalization of $\mathbf{M}$. \cite{banerjee2021074105}
For this reason, the PE-IP-ADC approach presents an attractive alternative to PE-EE-ADC for the direct and efficient calculations of ionized states in realistic environments.

\subsection{Benchmark}
\label{sec:results:benchmark}

\begin{figure*}[t!]
	\includegraphics[width=0.8\textwidth]{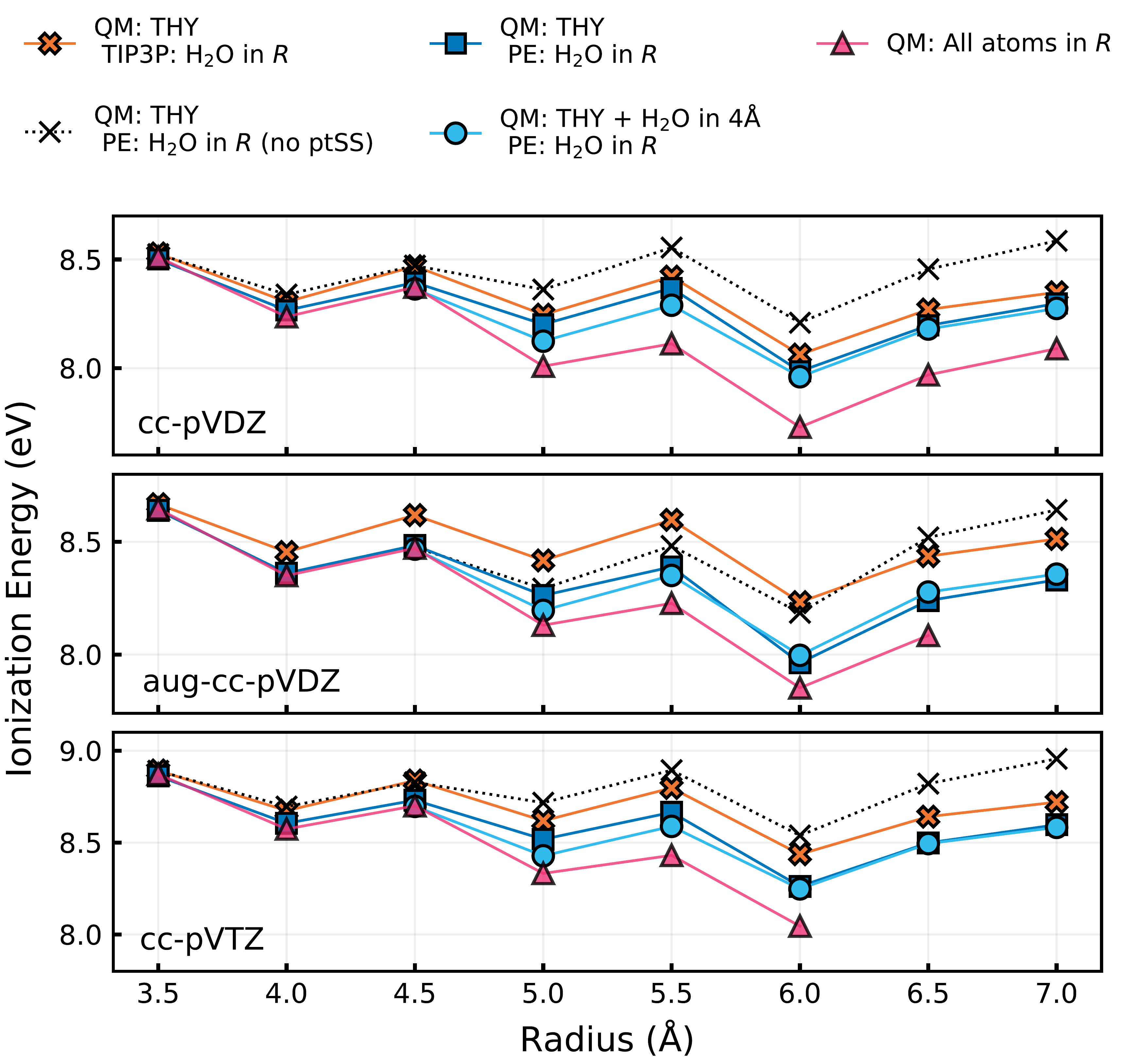}	
	\captionsetup{justification=raggedright,singlelinecheck=false}
	\caption{
		First ionization energy (eV) of thymine microsolvated in water plotted as a function of spherical solvation shell radius ($\angstrom$) relative to the thymine center of mass. 
		The results were computed using IP-ADC(2) with three basis sets for five QM/environment models: 
		(i) QM: all molecules, environment: none (triangles);
		(ii) QM: thymine + \ce{H2O} in 4 $\angstrom$ radius, environment: remaining \ce{H2O} with PE (circles);
		(iii) QM: thymine, environment: all \ce{H2O} with PE (squares);		
		(iv) QM: thymine, environment: all \ce{H2O} with PE without $\Delta E^{\mathrm{ptSS}}_{0\rightarrow n}$ (regular crosses);
		(v) QM: thymine, environment: all \ce{H2O} with TIP3P (bold crosses).
	}
	\label{fig:adc2}
\end{figure*}

\begin{figure*}[t!]
	\includegraphics[width=.8\textwidth]{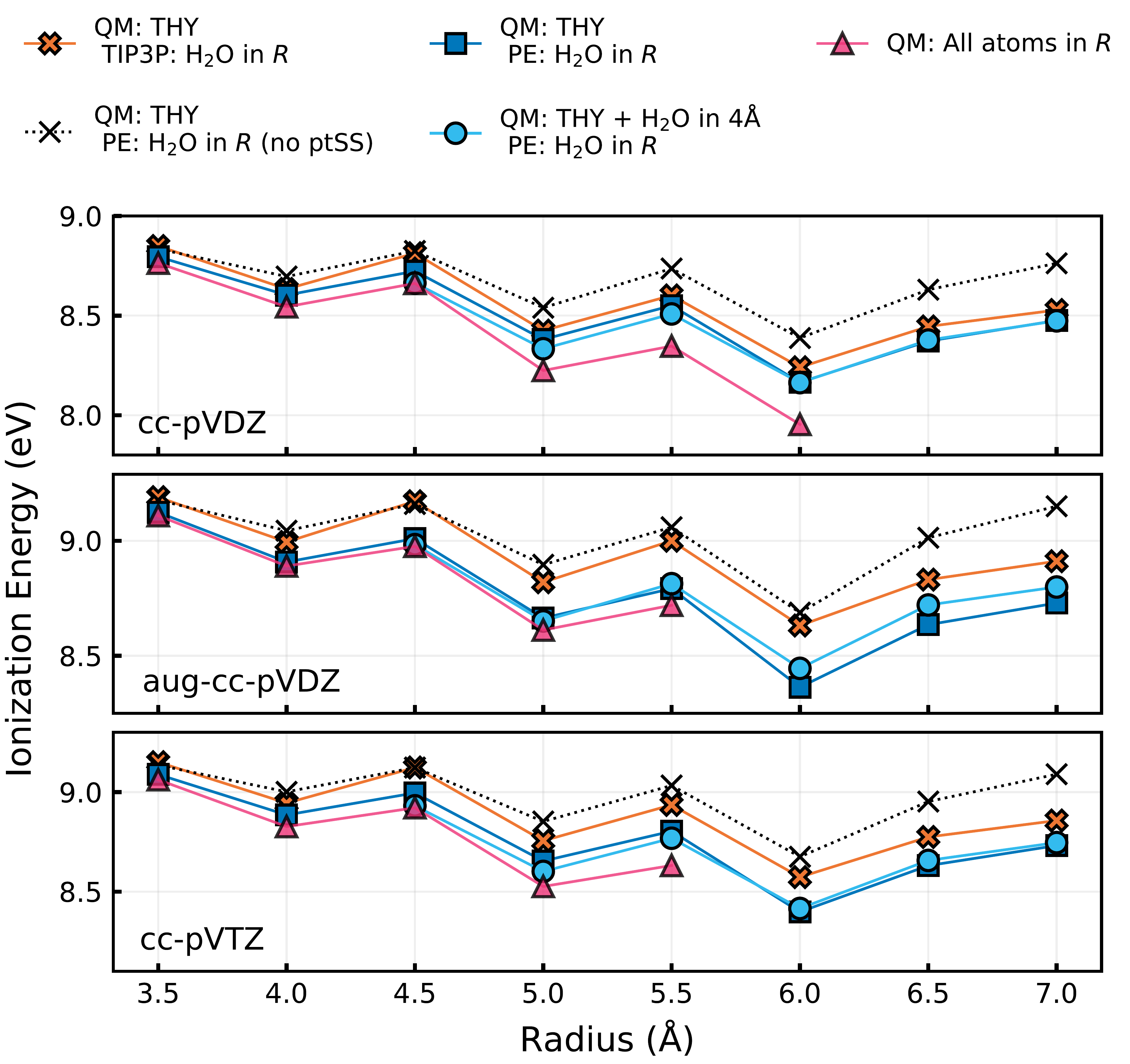}	
	\captionsetup{justification=raggedright,singlelinecheck=false}
	\caption{
		First ionization energy (eV) of thymine microsolvated in water plotted as a function of spherical solvation shell radius ($\angstrom$) relative to the thymine center of mass. 
		The results were computed using IP-ADC(3) with three basis sets for five QM/environment models: 
		(i) QM: all molecules, environment: none (triangles);
		(ii) QM: thymine + \ce{H2O} in 4 $\angstrom$ radius, environment: remaining \ce{H2O} with PE (circles);
		(iii) QM: thymine, environment: all \ce{H2O} with PE (squares);		
		(iv) QM: thymine, environment: all \ce{H2O} with PE without $\Delta E^{\mathrm{ptSS}}_{0\rightarrow n}$ (regular crosses);
		(v) QM: thymine, environment: all \ce{H2O} with TIP3P (bold crosses).
	}
	\label{fig:adc3}
\end{figure*}

As evidenced by the large $\Delta E^{\mathrm{ptSS}}_{0\rightarrow n}$ perturbative corrections calculated in \cref{sec:results:validation}, the ionized QM region interacts quite strongly with the environment.
Here, we assess the accuracy of polarizable embedding for such strong QM--environment interactions by comparing the PE-IP-ADC and all-quantum IP-ADC ionization energies for the spherical thymine + water clusters shown in \cref{fig:solv_radius}. 
\cref{fig:adc2,fig:adc3} show the IP-ADC($n$) results ($n$ = 2 and 3, respectively) with and without polarizable embedding for the clusters with radii ($R$) ranging from 3.5 to 7.0 $\angstrom$ and three basis sets (cc-pVDZ, aug-cc-pVDZ, and cc-pVTZ). 
The ionization energies were calculated using five QM/environment models, namely: 
(i) QM: all molecules, environment: none (all-quantum IP-ADC($n$));
(ii) QM: thymine + water molecules within 4 $\angstrom$ radius, environment: remaining waters described with PE;
(iii) QM: thymine, environment: all waters described with PE;		
(iv) QM: thymine, environment: all waters described with PE without the $\Delta E^{\mathrm{ptSS}}_{0\rightarrow n}$ correction;
(v) QM: thymine, environment: all waters described with TIP3P (MM embedding).

The best agreement with the all-quantum IP-ADC results is shown by the PE-IP-ADC models with and without water molecules in the QM region.
At the second order in perturbation theory (\cref{fig:adc2}), the error in PE-IP-ADC(2) ionization energy relative to IP-ADC(2) reaches its maximum of $\sim$ 0.2 to 0.3 eV for $R$ = 5.5 $\angstrom$ and remains relatively constant for solvation shells with larger radii.
Incorporating water molecules in the QM region has a very small effect on ionization energy for most calculations.
Increasing the basis set from cc-pVDZ to aug-cc-pVDZ and cc-pVTZ reduces the polarizable embedding error to $\sim$ 0.2 eV.
The IP-ADC(2) calculations with the MM (TIP3P) environment show significantly larger errors ($\sim$ 0.4 to 0.5 eV) compared to PE-IP-ADC(2), indicating the importance of polarization effects.
The $\Delta E^{\mathrm{ptSS}}_{0\rightarrow n}$ correction is crucially important in the PE-IP-ADC calculations providing a large ($\sim$ 0.4 to 0.5 eV) contribution to the ionization energy of PE-IP-ADC(2). 

Similar results are obtained for the third-order methods (\cref{fig:adc3}).
Interestingly, we find that the polarizable embedding error of PE-IP-ADC(3) is somewhat smaller than that of PE-IP-ADC(2) for all three considered basis sets. 
Using the aug-cc-pVDZ basis set, the  PE-IP-ADC(3) error in ionization energy for $R$ =  5.5 $\angstrom$ is $\sim$ 0.1 eV.
Due to the higher cost of IP-ADC(3) calculations, we cannot confirm if this error remains the same for larger $R$.
As for the ADC(2) results, increasing the basis set from aug-cc-pVDZ to cc-pVTZ has a minor effect on computed ionization energies.

Overall, our benchmark results indicate that the PE-IP-ADC approach delivers accurate ionization energies of solvated thymine clusters with errors ranging from 0.1 to 0.2 eV when using the aug-cc-pVDZ or cc-pVTZ basis sets.
Importantly, the PE-IP-ADC calculations can be performed at a small fraction of IP-ADC computational cost, as demonstrated by the comparison of timings in the {Supplementary Information}.
To showcase the power of PE-IP-ADC, in \cref{sec:results:solvated_thymine} we use this approach to calculate the vertical ionization energy of thymine solvated in bulk water. 

\subsection{Ionization Energy of Aqueous Thymine}
\label{sec:results:solvated_thymine}

\begin{figure*}[t!]
	\includegraphics[width=.7\textwidth]{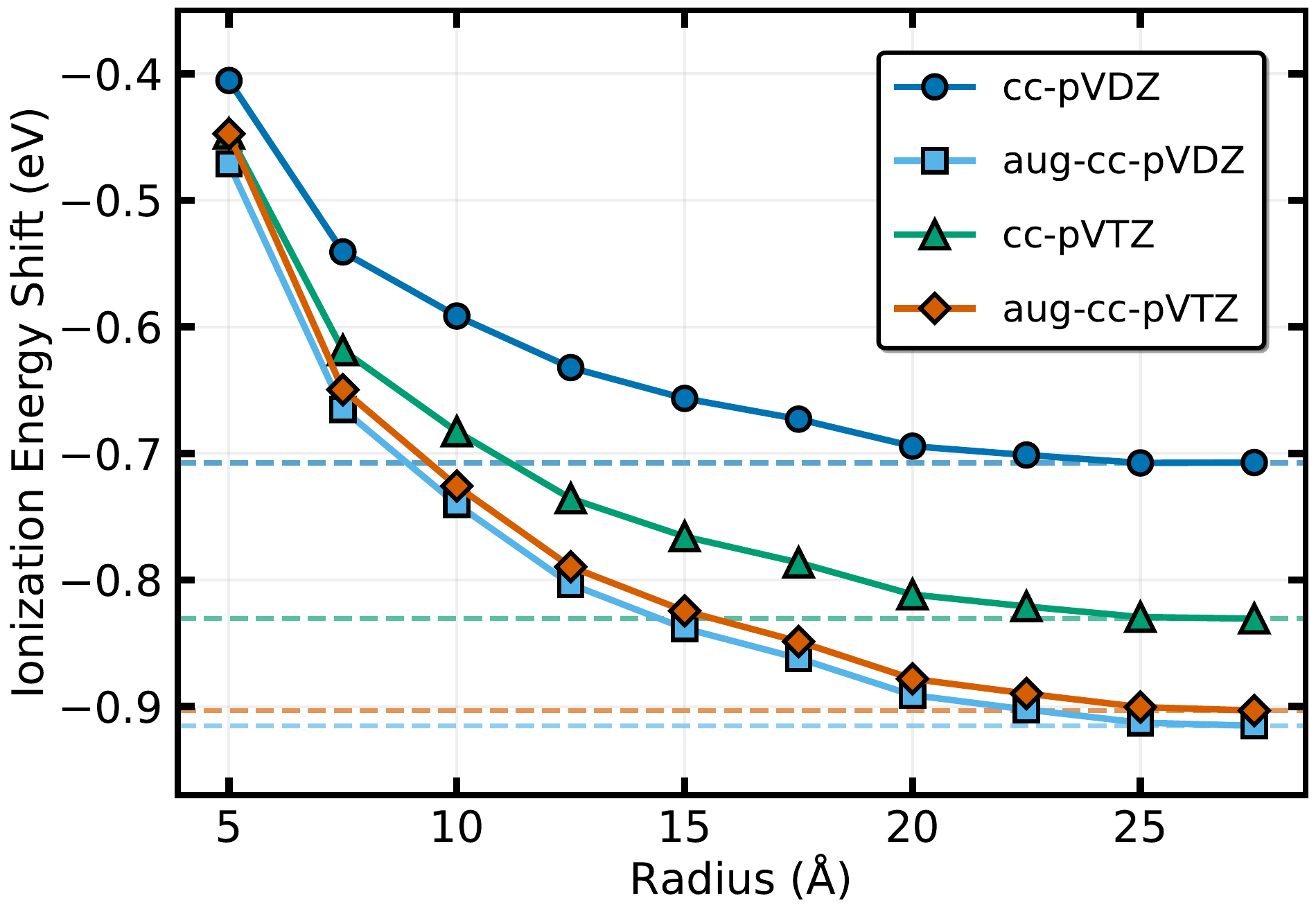}	
	\captionsetup{justification=raggedright,singlelinecheck=false}
	\caption{
		Solvent-induced shift in vertical ionization energy of aqueous thymine computed using PE-IP-ADC(2) with four basis sets as a function of solvation shell radius. 
		See \cref{sec:comp} for computational details.
	}
	\label{fig:basis_sets}
\end{figure*}

\begin{figure*}[t!]
	\includegraphics[width=1\textwidth]{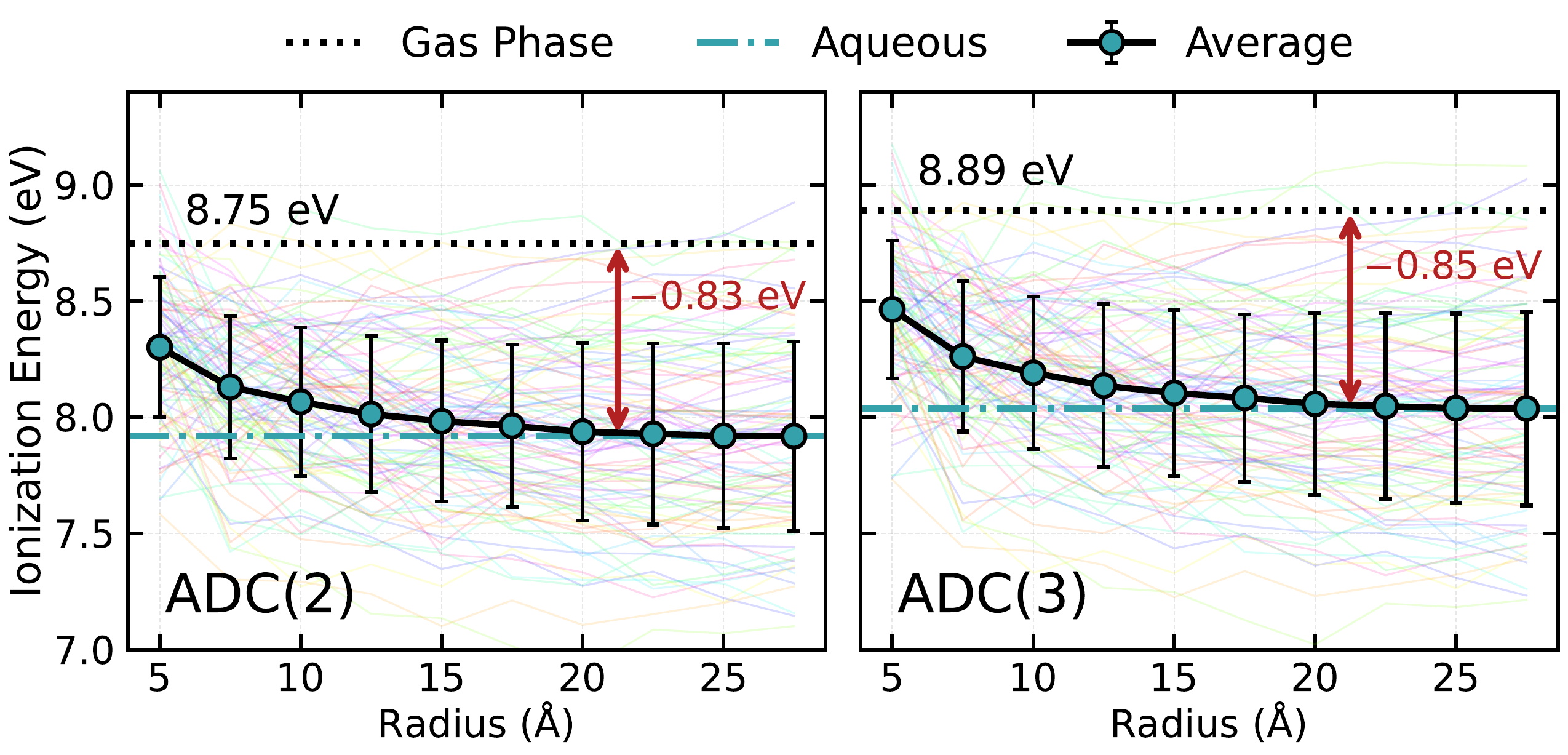}  
	\caption{
		Vertical ionization energy (VIE, eV) of aqueous thymine computed as a function of solvation shell radius using PE-IP-ADC(2) and PE-IP-ADC(3) with the cc-pVTZ basis set. 
		Results are averaged over 100 geometries sampled during a molecular dynamics simulation (\cref{sec:comp}), with each faint line corresponding to one sample.
		Vertical bars indicate the spread (one standard deviation) of VIE due to geometry fluctuations. 
		Also indicated is the gas-phase VIE of thymine computed at the IP-ADC(2) and IP-ADC(3) levels of theory, respectively.
	}
	\label{fig:md_avg}
\end{figure*}

\begin{table*}[t!]
	\centering
	\caption{Vertical ionization energy (VIE, eV) of thymine in the gas phase and bulk water calculated using several levels of theory in comparison to experimental results.\cite{slavicek20096460-6467}
		The theoretical VIE were averaged over 100 geometries from a molecular dynamics simulation (see \cref{sec:comp} for details).
		Also shown are the solvent-induced shifts in averaged VIE ($\Delta$VIE, eV) and the standard deviations in VIE due to geometry fluctuations ($\pm$, eV).
	}
	\centering
	\footnotesize
	\setstretch{1}
	\begin{threeparttable}
		\begin{tabular}{lc c c c}
			\midrule
			{Method}  & {Gas-phase VIE} & {Aqueous VIE} & $\Delta$VIE \\ 
			\midrule
			PE-IP-ADC(2)/cc-pVDZ  & 8.40 & 7.69 $\pm$ 0.41 & $-$0.71 \\
			PE-IP-ADC(2)/aug-cc-pVDZ & 8.53 & 7.61 $\pm$ 0.41 & $-$0.92 \\ 
			PE-IP-ADC(2)/cc-pVTZ & 8.75  &  7.92 $\pm$ 0.41 & $-$0.83  \\
			PE-IP-ADC(2)/aug-cc-pVTZ & 8.82  &  7.91 $\pm$ 0.41 & $-$0.91  \\
			\midrule 
			PE-IP-ADC(3)/cc-pVTZ &  8.89 & 8.04 $\pm$ 0.42 &   $-$0.85 \\
			PE-IP-ADC(3)/aug-cc-pVTZ (est.) & 9.01 & 8.08 $\pm$ 0.42\tnote{a} &   $-$0.93\tnote{a} \\
			\midrule 
			IP-EOM-CCSD/EFP/6-31+G(d)\cite{ghosh20116028-6038} & 8.97 & 8.07 & $-$0.90 \\
			IP-EOM-CCSD/EFP/cc-pVTZ (est.)\cite{ghosh20116028-6038} & 9.14 & 8.24\tnote{b} & $-$0.90\tnote{b} \\
			Experiment\cite{slavicek20096460-6467} & 9.20 & 8.30 & $-$0.90 \\
			\hline \hline 
		\end{tabular}
		\begin{tablenotes}
			\item[a] The PE-IP-ADC(3)/aug-cc-pVTZ $\Delta$VIE was estimated as $\Delta$VIE = $\Delta$VIE[ADC(3)/cc-pVTZ] + $\Delta$VIE[ADC(2)/aug-cc-pVTZ] $-$ $\Delta$VIE[ADC(2)/cc-pVTZ]. The resulting $\Delta$VIE was used to estimate the aqueous VIE at the PE-IP-ADC(3)/aug-cc-pVTZ level of theory. 
			\item[b] Calculated using the gas-phase VIE from IP-EOM-CCSD/cc-pVTZ  and $\Delta$VIE from IP-EOM-CCSD/EFP/6-31+G(d).
		\end{tablenotes}
	\end{threeparttable}
	\label{tab:final}
\end{table*}

Understanding the stability of DNA under ionizing radiation or in the presence of strong oxidants requires accurately quantifying the ionization energies of its nucleobases in condensed phase environments.\cite{dekker200129-33, hutchinson1985115-154}
Among the four DNA nucleobases, thymine is known to have the highest ionization energy that is strongly influenced by solvent and physiological conditions.\cite{khistyaev2011313, bravaya201012305-12317}
Experimentally, the VIE of thymine in water is estimated from the photoelectron measurements of aqueous deoxythymidine (thymine bonded to a deoxyribose sugar) where ionizing the thymine moiety requires 8.3 eV.\cite{slavicek20096460-6467} 
Comparing to the VIE of thymine in the gas phase (9.2 eV), the experimental solvent-induced shift in thymine VIE ($\Delta$VIE) is estimated to be $-$0.9 eV.
Several computational studies of the solvation effects on thymine VIE have been reported.\cite{slavicek20096460-6467, pluharova20151209-1217a, khistyaev2011313, close20059279-9283, close20084405-4409}
The best theoretical estimate of thymine VIE in bulk water was computed by Ghosh et al.\cite{ghosh20116028-6038} using equation-of-motion coupled cluster theory with single and double excitations (IP-EOM-CCSD) combined with the effective fragment potential model (EFP) to describe the solvent environment.
Using the 6-31+G(d) basis set and sampling over 100 configurations during the 10 ns molecular dynamics simulation, $\Delta$VIE was computed to be $-$0.9 eV, in a perfect agreement with the experimental measurements.\cite{ghosh20116028-6038}

Here, we use the PE-IP-ADC(2) and PE-IP-ADC(3) methods to calculate the VIE of thymine in bulk water. 
Our computational approach is described in detail in \cref{sec:comp}.
In short, we carried out a 10 ns molecular dynamics simulation of thymine in a box of water and extracted 100 geometries at equal time intervals.
To study the effect of solvent environment on VIE, at each geometry the PE-IP-ADC calculations were performed for thymine and all water molecules within a sphere with radius $R$ (from 5 to 27.5 $\angstrom$) relative to the thymine center of mass.
As in \cref{sec:results:benchmark}, thymine and water molecules within the  4 $\angstrom$ radius were included in the QM region. 
All remaining molecules were described using polarizable embedding.

First, we investigate how basis set and solvation shell radius  ($R$) influence the VIE of aqueous thymine.
\cref{fig:basis_sets} shows the shift in thymine VIE ($\Delta$VIE) as a function of $R$ computed using PE-IP-ADC(2) for four basis sets: cc-pVDZ, aug-cc-pVDZ, cc-pVTZ, and aug-cc-pVTZ.
Relative to the gas-phase VIE calculated using IP-ADC(2) with the same basis set, the calculated VIE shifts  are fully converged with respect to the solvation shell size at $R$ = 25 $\angstrom$.
Using the cc-pVDZ basis set results in $\Delta$VIE of $-$0.71 eV.
Increasing the basis set to cc-pVTZ lowers $\Delta$VIE to $-$0.83 eV.
The aug-cc-pVDZ and aug-cc-pVTZ basis sets yield very similar $\Delta$VIE of $-$0.92 eV, highlighting the importance of diffuse basis functions.
We note that the QM/MM calculations with diffuse basis sets must be carried out with caution as they may suffer from electrons spilling out of the QM region.\cite{marefatkhah20201373-1381}
While we are unable to determine if our calculations are affected by electron spill-out, the convergence of results with respect to the basis set observed in \cref{fig:basis_sets} is quite normal, which indicates that this effect is likely to be small if present.

Next, we investigate the role of electron correlation by comparing the PE-IP-ADC(2) and PE-IP-ADC(3) results in \cref{fig:md_avg} calculated using the cc-pVTZ basis set.
Incorporating the third-order correlation effects in IP-ADC(3) increases the gas-phase VIE of thymine from 8.75 eV (IP-ADC(2)) to 8.89 eV, improving the agreement with the experimental gas-phase measurement of 9.20 eV.\cite{slavicek20096460-6467}
Consistent with the results in \cref{fig:basis_sets}, the PE-IP-ADC(3) VIE of aqueous thymine is fully converged with respect to the size of solvation shell at $R$ = 25 $\angstrom$ when averaged over 100 sampled geometries.
The resulting aqueous VIE of 8.04 eV corresponds to the PE-IP-ADC(3)/cc-pVTZ solvent-induced shift $\Delta$VIE = $-$0.85 eV, which is only $-$0.02 eV lower compared to the $\Delta$VIE at the PE-IP-ADC(2)/cc-pVTZ level of theory.
In addition to the averaged VIE, \cref{fig:md_avg} shows the instantaneous VIE computed at different geometries during a molecular dynamics simulation.
We note that the spread in VIE distribution represents the changes in VIE due to geometry fluctuations and are not indicative of the errors in statistical sampling. 
The calculated standard deviations for the PE-IP-ADC(2) and PE-IP-ADC(3) VIE distributions at $R$ = 27.5 $\angstrom$ are 0.41 and 0.42 eV, respectively, in a close agreement with the results from Ref.\@ \citenum{ghosh20116028-6038}.

\cref{tab:final} summarizes the results of our PE-IP-ADC calculations and compares them to the IP-EOM-CCSD/EFP data from Ghosh et al.\cite{ghosh20116028-6038} and experimental measurements.\cite{slavicek20096460-6467}  
The IP-ADC(3)/aug-cc-pVTZ VIE of gas-phase thymine was calculated to be 9.01 eV, in a good agreement with the experimental gas-phase VIE of 9.2 eV.
The PE-IP-ADC(2)/aug-cc-pVTZ and PE-IP-ADC(2)/cc-pVTZ results were used to estimate the $\Delta$VIE and aqueous VIE of thymine at the PE-IP-ADC(3)/aug-cc-pVTZ level of theory (see \cref{tab:final} for details).
The estimated $\Delta$VIE = $-$0.93 eV is in an excellent agreement with the experimental measurement and IP-EOM-CCSD/EFP calculations  ($-$0.90 eV).
Notably, the lower computational cost of PE-IP-ADC relative to IP-EOM-CCSD allows to perform calculations with significantly larger basis sets and QM regions while accurately incorporating the strong QM--environment polarization effects.  

\section{Conclusions}
\label{sec:conclusions}

In this work, we developed an efficient and accurate approach for simulating ionized electronic states in realistic chemical environments based on algebraic diagrammatic construction theory with polarizable embedding (PE-IP-ADC). 
Our benchmark results demonstrate that PE-IP-ADC provides accurate description of strong interactions between the ionized states and their environment with errors in ionization energies of $\sim$ 0.1 to 0.3 eV.
We demonstrated the capabilities of second- and third-order PE-IP-ADC methods (PE-IP-ADC($n$), $n$ = 2, 3) by calculating the vertical ionization energy (VIE) of thymine in bulk water.
Sampling over 100 configurations during a molecular dynamics simulation and using large basis sets (up to aug-cc-pVTZ), PE-IP-ADC(2) and PE-IP-ADC(3) predict the solvent-induced shift in thymine VIE of $-$0.92 and $-$0.93 eV, respectively, in an excellent agreement with experimental\cite{slavicek20096460-6467} and high-level theoretical\cite{ghosh20116028-6038} results ($-$0.9 eV).
The PE-IP-ADC(3) method shows the best agreement with experiment for the gas-phase and aqueous VIE (9.2 and 8.3 eV), underestimating both of their values by $\sim$ 0.2 eV.

Our work builds upon the previous development of ADC with PE,\cite{scheurer20184870-4883} extending it to the simulations of charged excitations.
Although the results presented in this study are encouraging, the accuracy of perturbative correction employed in our PE-IP-ADC implementation needs to be assessed for a broader range of chemical systems.
The PE-IP-ADC methods can be further improved by introducing the self-consistent treatment of environment effects in the effective Hamiltonian, improving the level of electron correlation treatment, \cite{leitner20247680-7690, sokolov2018204113a,chatterjee20195908-5924,chatterjee20206343-6357,mazin20216152-6165} and using better embedding models.\cite{olsen20155344-5355,vandenheuvel20233248-3256} 
Applications of PE-IP-ADC to larger QM regions would require lowering their computational cost by using local correlation,\cite{schutz2001661-681, werner20038149-8160a, riplinger2013034106} frozen natural orbital,\cite{taube2005837-850, mester2018094111, mukhopadhyay2023084113} or tensor factorization techniques.\cite{hohenstein2012044103, parrish2012224106, hohenstein2012221101}
Finally, the PE-IP-ADC methods need to be integrated with  better molecular dynamics and sampling techniques, to enable accurate calculations of vibrational and thermodynamic properties.

\suppinfo
Computational wall times for the benchmark calculations in \cref{sec:results:benchmark}, average $\Delta E^{\mathrm{ptSS}}_{0\rightarrow n}$ and VIE data for the calculations in \cref{sec:results:solvated_thymine}, and the Cartesian coordinates for solvated thymine structures. 

\acknowledgement
The authors acknowledge the donors of the American Chemical Society Petroleum Research Fund for supporting this research (PRF Grant No.\@ 65903-ND6). 
We also thank Dr.\@ Terrence Stahl for the help with implementation and Prof.\@ John Herbert for insightful discussions. 
Computations were performed at the Ohio Supercomputer Center under Project No.\@ PAS1963.\cite{OhioSupercomputerCenter1987}

%\bibliography{james_ref}

\providecommand{\latin}[1]{#1}
\makeatletter
\providecommand{\doi}
{\begingroup\let\do\@makeother\dospecials
	\catcode`\{=1 \catcode`\}=2 \doi@aux}
\providecommand{\doi@aux}[1]{\endgroup\texttt{#1}}
\makeatother
\providecommand*\mcitethebibliography{\thebibliography}
\csname @ifundefined\endcsname{endmcitethebibliography}
{\let\endmcitethebibliography\endthebibliography}{}

\end{document}